\documentclass{article}
\usepackage{emulateapj}
\usepackage{amsmath}
\usepackage{amssym}
\usepackage{psfig}
\usepackage{times}

\newcommand{\gta}{\gtrsim}
\newcommand{\lta}{\lesssim}

\newcommand{\eV}{\mbox{e\kern-0.35mm{V}}}
\newcommand{\keV}{\mbox{ke\kern-0.35mm{V}}}

\newcommand{\Tbb}{\ensuremath{T_{\mbox{\scriptsize\sc bb}}}}
\newcommand{\Tc}{\ensuremath{T_{\mbox{\scriptsize\sc c}}}}

\newcommand{\Tmax}{\ensuremath{T_{\mbox{\scriptsize\sc max}}}}
\newcommand{\lc}{\ensuremath{l_{\mbox{\scriptsize\sc c}}}}

\newcommand{\sigmaT}{\ensuremath{\sigma_{\mbox{\scriptsize T}}}}

\newcommand{\sigmasb}{\ensuremath{\sigma_{\mbox{\scriptsize sb}}}}
\newcommand{\tauT}{\ensuremath{\tau_{\mbox{\scriptsize T}}}}
\newcommand{\taue}{\ensuremath{\tau_{\mbox{\scriptsize e}}}}
\newcommand{\taup}{\ensuremath{\tau_{\mbox{\scriptsize p}}}}
\newcommand{\R}{\ensuremath{R}}
\newcommand{\Rc}{\ensuremath{R_{\mbox{\scriptsize c}}}}
\newcommand{\Rd}{\ensuremath{R_{\mbox{\scriptsize d}}}}
\newcommand{\Rs}{\ensuremath{R_{\mbox{\scriptsize s}}}}
\newcommand{\fc}{\ensuremath{f_{\mbox{\scriptsize c}}}}
\newcommand{\Lc}{\ensuremath{L_{\mbox{\scriptsize c}}}}

\newcommand{\ld}{\ensuremath{l_{\mbox{\scriptsize d}}}}
\newcommand{\me}{\ensuremath{m_{\mbox{\scriptsize e}}}}
\newcommand{\np}{\ensuremath{n_{\mbox{\scriptsize p}}}}
\newcommand{\nee}{\ensuremath{n_{\mbox{\scriptsize e}}}}
\newcommand{\fdd}{\ensuremath{f_{\rm d}}}

\newcommand{\diff}{\ensuremath{{\rm d}}}

\newcommand{\Msun}{\ensuremath{{\rm M}_\odot}}

\newcommand{\Te}{\Tc}

\newcommand{\avg}[1]{\ensuremath{\langle#1\rangle}}

\makeatletter
\@ifundefined{chapter}{\def\thebibliography#1{
\section*{References}
  \list
  {\relax}{\setlength{\labelsep}{0em}
        \setlength{\itemindent}{-\bibhang}
        \setlength{\itemsep}{\parskip}
        \setlength{\parsep}{0pt}
        \setlength{\leftmargin}{\bibhang}}
    \def\newblock{\hskip .11em plus .33em minus .07em}
    \sloppy\clubpenalty4000\widowpenalty4000
    \sfcode`\.=1000\relax}}%
{\def\thebibliography#1{
  \list
  {\relax}{\setlength{\labelsep}{0em}
        \setlength{\itemindent}{-\bibhang}
        \setlength{\itemsep}{\parskip}
        \setlength{\parsep}{0pt}
        \setlength{\leftmargin}{\bibhang}}
    \def\newblock{\hskip .11em plus .33em minus .07em}
    \sloppy\clubpenalty4000\widowpenalty4000
    \sfcode`\.=1000\relax}}

\newlength{\bibhang}
\let\@internalcite\cite
\def\cite{\@ifstar{\citey}{\citefull}}
\def\citefull{\def\astroncite##1##2{##1\ ##2}\@internalcite}
\def\citey{\def\astroncite##1##2{##1\ (##2)}\@internalcite}
\def\citeyear{\def\astroncite##1##2{##2}\@internalcite}
\def\citename{\def\astroncite##1##2{##1}\@internalcite}
\def\@citex[#1]#2{\if@filesw\immediate\write\@auxout{\string\citation{#2}}\fi
  \def\@citea{}\@cite{\@for\@citeb:=#2\do
    {\@citea\def\@citea{; }\@ifundefined
       {b@\@citeb}{{\bf ??}\@warning
       {Citation `\@citeb' on page \thepage \space undefined}}%
{\csname b@\@citeb\endcsname}}}{#1}}

\def\@cite#1#2{#1\if@tempswa #2\fi} 
\def\@biblabel#1{}

\def\astroncite#1#2{#1\ #2}
\makeatother

\pssilent

\begin{document}

\title{Self-Consistent Thermal 
Accretion Disk Corona Models for Compact Objects II. Application
to Cygnus X-1}

\author{James B. Dove\altaffilmark{1,2}, J\"orn Wilms\altaffilmark{3,1},
Michael Maisack\altaffilmark{3}, and Mitchell
C. Begelman\altaffilmark{1,2}}

\altaffiltext{1}{JILA, University of Colorado and National Institute of
Standards and Technology, Campus Box 440, Boulder, CO
80309-0440.\\ \{dove,mitch\}@rocinante.colorado.edu}

\altaffiltext{2}{Department of Astrophysical, Planetary, 
and Atmospheric Sciences, University of
Colorado, Boulder, Boulder, CO 80309-0391.}

\altaffiltext{3}{Institut f\"ur Astronomie und Astrophysik,
Abt.~Astronomie, Waldh\"auser Str. 64, D-72076 T\"ubingen, Germany.\\
\{wilms,maisack\}@astro.uni-tuebingen.de}

\begin{abstract}
We apply our self-consistent accretion disk corona (ADC) model, with two
different geometries, to the broad-band X-ray spectrum of the black hole
candidate Cygnus X-1. As shown in a companion paper (Dove, Wilms, and
Begelman), models where the Comptonizing medium is a slab surrounding the
cold accretion disk cannot have a temperature higher than about 120\,\keV\
for optical depths greater than 0.2, resulting in spectra that are much
softer than the observed 10--30\,\keV\ spectrum of Cyg X-1. In addition,
the slab geometry models predict a substantial ``soft excess'' at low
energies, a feature not observed for Cyg~X-1, and Fe K$\alpha$ fluorescence
lines that are stronger than observed.  Previous Comptonization models in
the literature invoke a slab geometry with the optical depth $\tauT \gta
0.3$ and the coronal temperature $\Tc \sim 150$ \keV, but they are not
self-consistent.  Therefore, ADC models with a slab geometry are not
appropriate for explaining the X-ray spectrum of Cyg~X-1.  Models with a
spherical corona and an exterior disk, however, predict much higher
self-consistent coronal temperatures than the slab geometry models.  The
higher coronal temperatures are due to the lower amount of reprocessing of
coronal radiation in the accretion disk, giving rise to a lower Compton
cooling rate.  Therefore, for the sphere$+$disk geometry, the predicted
spectrum can be hard enough to describe the observed X-ray continuum of
Cyg~X-1 while predicting Fe fluorescence lines having an equivalent width
of $\sim 40$\,\eV.  Our best-fit parameter values for the sphere$+$disk
geometry are $\tauT \approx 1.5$ and $\Tc \approx 90$\,\keV.
\end{abstract}

\section{Introduction}
Due to its high X-ray brightness and its apparent large mass, Cygnus~X-1
is the most studied Galactic black-hole candidate (BHC).  Although it is
almost certain that the production of the high-energy radiation is
associated with the accretion of matter, the exact details are uncertain.
The source of accreting matter is most probably a combination of Roche-lobe
overflow and accretion out of the stellar wind of the companion, the
O9.7Iab supergiant HDE~226868 (\cite{gies86b}).  Recent mass determinations
give a mass of about 18\,\Msun\ for the companion and 10\,\Msun\ for the
compact object (\cite{herrero95a}, in agreement with
\cite{dolan92a,sokolov87a,bochkarev86a,aab84a,hutchings78a} and references
therein), making the latter a good candidate for being a black hole.

Cyg~X-1 is usually observed in the so-called ``X-ray low, $\gamma$-ray
high'' state, which is roughly described by a power-law with a photon index
$\alpha\approx 1.7$, modified by an exponential cut-off with an $e$-fold
energy of about 100\,\keV\ (see \cite{oda77a,liang84,tanaka95a} for
detailed reviews of the observations).  The most popular model used
for explaining the high-energy spectrum of Cyg~X-1 (and other BHCs), while
in the ``low state,'' involves inverse Comptonization of soft photons by a
hot plasma, usually referred to as a corona. While the Comptonization model
by \cite*{sunyaev80} has been used widely in the past (e.g.,\
\cite{grebenev93,doeber95a}), \cite*{titarchuk94b} showed that the typical
fit-parameters found for Cyg~X-1 are outside the model's range of validity
(and cannot be used for interpreting the physical conditions in the
Comptonizing medium), and he developed a more generalized theory.  Using a
composite spectrum from EXOSAT, GRANAT, and OSSE observations,
\cite*{titarchuk94b} found that this new accretion
disk corona (ADC) model, having an optical depth
$\taue\approx 0.6$, a coronal temperature $k\Te\approx 150\,\keV$, and a
slab geometry allowed for a {\em rough} description of Cyg~X-1.

Although Comptonization apparently accounts for the general behavior of the
observed X-ray spectrum of Cyg~X-1, it appears that a reprocessing
component is needed for the full description of the spectrum.  Formal
descriptions of the different observations in terms of a reflection
component with an underlying Comptonization spectrum give better fits to
the data than do pure Comptonization models.  The presence of a weak
fluorescence line from neutral iron
(\cite{kitamoto90a,done92a,marshall93a,ebisawa96b}) and the deviation of
the 2--60\,\keV-band from a pure power law seen by HEAO~1-A2
(\cite{inoue89a,done92a,gierlinski97b}), which is usually interpreted as a
Compton reflection hump, both indicate the presence of cold or slightly
ionized material present in the source, in addition to the Comptonizing
medium.  Therefore, since the radiation processes of Comptonization and
reprocessing of coronal radiation in the cold accretion disk cover a large
range of photon energies ($\lta 2$\,\keV\ to $\sim 250$\,\keV) in the
predicted X-ray spectrum, it is necessary to use-broad-band spectral
observations to place meaningful constraints on the models.

As we discussed in our companion paper (Dove, Wilms, \& Begelman 1997,
hereafter paper~I), one serious drawback to solving the radiation transfer
problem for an ADC with analytic and ``linear'' Monte Carlo techniques
(where each photon particle is propagated through a fixed background) is
that the physical properties of the corona must be defined {\it a priori}.
The coronal properties and the radiation field are strongly coupled through
the processes of Compton cooling, Compton upscattering, pair production and
annihilation.  Therefore, it is not clear whether the {\em assumed} coronal
properties are self-consistent with the resulting radiation field.

We have recently developed an ADC model in which the radiation field, the
temperature and opacity of the corona, and the reprocessing of coronal
radiation in the accretion disk are calculated self-consistently (paper~I).
We find that all previous ADC models having a slab geometry and used to
describe the X-ray spectrum of Cyg~X-1 are not self-consistent, since the
coronal temperatures and opacities are not within the physically allowed
range of combinations of these parameters (see Fig.~2 of paper~I).
\cite*{haardt93a} claim that their ``best-fit'' model, having a coronal
temperature $k\Tc = 150$\,\keV\ and an optical depth $\tauT = 0.3$, agree
within the 3$\sigma$ confidence level of their allowed $k\Tc - \tauT$
relation.  However, due to the quality of the data, the 3$\sigma$ contour
covers a very large region of the $k\Tc-\tauT$ parameter space, of which
only a very small part is within the self-consistently allowed region.

In this paper, we apply our self-consistent model to Cyg~X-1, using a more
complete composite spectrum than was available to previous work
(\cite{wilms97a}).  We study two geometries: (1) the standard slab geometry
and (2) a spherical corona with an exterior accretion disk.  At the time of
writing, simultaneous X-ray spectra covering the range from 2\,\keV\ to
250\,\keV\ were not generally available.  We therefore had to use a
non-simultaneous composite spectrum from several instruments to cover this
energy range (each observation was taken while Cyg~X-1 was in its ``low''
state). A detailed description of the individual data comprising the
composite spectrum and a preliminary analysis are given in
\cite*{wilms97a}.  \footnote{We note that, due to a bug in the treatment of
reprocessing of coronal radiation, the conclusion made by us in
\cite*{wilms97a} that the slab geometry model is able to describe the
broad-band spectrum of Cyg~X-1 is erroneous.  As we show below, the slab
geometry model predicts a spectrum that is much softer than the observed
spectrum. Due to the bug, the yields for Compton reflection/fluorescence
and thermalization were reversed, causing $\lta 20$\% (instead of $\gta
80$\%) of the coronal radiation to be reprocessed into thermal radiation
and leading to a corresponding erroneous Compton cooling rate. As discussed
in paper~I, the current version of our code has been successfully compared
to other reprocessing models (e.g., \cite{george90}).}

\section{The Numerical Code}\label{sec:models}
It appears impossible to produce an analytical model of an ADC
that properly accounts for the self-consistent thermal and opacity
distributions of the corona while solving correctly the radiative transfer
problem of an angle-dependent, high-energy radiation field in a
semi-relativistic, non-uniform plasma (including reprocessing of radiation
in the accretion disk). Therefore, numerical methods are needed.  For the
computations presented here, we use a non-linear Monte Carlo (NLMC) code to
calculate self-consistently the temperature and opacity of the corona as
well as the radiation field within and external to the corona.  For the two
geometries considered here, Compton scattering, photon-photon pair
production, and pair annihilations are taken into account within the
corona. For the reprocessing of radiation within the accretion disk,
Compton scattering and photo-absorption, resulting in fluorescent line
emission and thermal emission, are considered.  We enforce charge
neutrality by requiring $n_{\rm e} - n_{+} = n_{\rm p}$, where $n_{\rm e}$,
$n_{+}$, and $n_{\rm p}$ are the number densities of electrons, positrons,
and protons, respectively.  For this paper, the spatial distribution of the
seed opacity (the opacity not including the contribution due to
electron-positron pairs) was assumed to be uniform.

The free parameters of our models are: (1) the seed opacity of the corona,
\taup, (2) the temperature of the accretion disk as a function of radius,
\Tbb$(R)$, and (3) the compactness parameter of the corona, \lc.
For a set of free parameters, the NLMC code is iterated until the
temperature of the corona, \Tc, the total opacity, $\tauT = \taup +
2\tau_+$, and the radiation field have reached a steady state.  Thermal
equilibrium is achieved when the {\em local} heating rate is balanced by
the Compton cooling rate. The pair opacity is determined by requiring
pair-annihilations to balance pair-production.  Once the system has reached
steady state, the spectra of escaping radiation are recorded (in 10 bins
of different inclination angles) until satisfactory statistics are
obtained. A detailed description of the NLMC code and its use for ADC is
given in paper~I.

\section{Slab Geometry}\label{sec:slabresults}
\subsection{The Model}
The most common geometry used to model the spectra of BHCs is the slab
(plane-parallel) geometry.  For this geometry, we assume that the accretion
disk corona is situated above and below an optically thick, geometrically
thin, cold accretion disk, that all corona and disk properties are
constant with respect to radius, and that azimuthal symmetry applies.  The
coronal compactness parameter is defined to be
\begin{equation}\label{eq:lcslab}
\lc = \frac{\sigmaT}{m_{\rm e}c^3}z_0\Psi_{\rm c}
= 0.7\left(\frac{\Lc}{0.1L_{\rm edd}}\right)\left(\frac{100 R_{\rm
s}}{\Rc}\right)\left(\frac{h}{0.1}\right),
\end{equation}
where $z_0$ is the scale-height of the corona, $L_{\rm c}$ is the
luminosity of the corona, \Rc\ is the radius of the corona, $L_{\rm edd}$ is
the Eddington luminosity, $\Rs$ is the Schwarzschild radius, $\Psi_{\rm c}$
is the rate of energy dissipation per unit area into the corona and $h =
z_0/\Rc$.  The specification of the physical mechanism heating the corona
is not necessary.  We simply assume that the energy is dissipated
uniformly with respect to height.  For a specific value of \lc, the
self-consistent coronal properties are degenerate with respect to $z_0$ and
$\Psi_{\rm c}$ (\cite{stern95a}).

As we discuss below, the predicted amount of thermal excess in the escaping
spectrum of radiation strongly depends on the temperature of the
accretion disk, \Tbb.  A lower limit to the temperature can be estimated by
assuming that the disk is heated solely by illumination from the corona.  In
this case, the temperature is approximated by balancing absorption
with thermal emission:
\begin{equation}\label{eq:illumslab}
\sigmasb \Tbb^4 = (1-A) F_{\rm c},
\end{equation}
where $F_{\rm c}$ is the flux of coronal radiation incident onto the
accretion disk (assumed here, by symmetry, to be equal to the flux of escaping
radiation), $\sigmasb$ is the Stefan-Boltzmann constant, and $A$ is the
angle-averaged albedo of the disk.  Using equation~(\ref{eq:lcslab}), the
disk temperature can be expressed as
\begin{multline}\label{eq:Tbbslab}
k\Tbb({\rm slab}) 
  = k\left(\frac{\me c^5}{\sigmaT\sigmasb G
M}\right)^{1/4}\left(\frac{\lc}{h}\right)^{1/4}\left(\frac{R_{\rm
s}}{\Rc}\right)^{1/4}\\ 
 \approx 220 \lc^{1/4}\left(\frac{10\Msun}{M}\right)^{1/4}\left(\frac{100
\Rs}{\Rc}\right)^{1/4}\left(\frac{0.1}{h}\right)^{1/4} {\rm eV}\\
 \approx 155 \left(\frac{10\Msun}{M}\right)^{1/4}\left(\frac{100
\Rs}{\Rc}\right)^{1/2}\left(\frac{\Lc}{0.1 L_{\rm edd}}\right)^{1/4} {\rm
eV}.
\end{multline}
We refer the reader to paper~I for a more detailed discussion of the slab
geometry models.

\subsection{Application to Cyg~X-1}\label{sec:slabcyg}
As discussed in paper~I, we find that the maximum self-consistent
temperature of the corona is $\Tmax \sim 120$\,\keV\ for total opacities
$\tauT \gta 0.2$, regardless of the compactness parameter \lc\ and \Tbb.
Once this maximum temperature, \Tmax, is reached, higher heating rates
result in a higher opacity due to pair production, causing more
reprocessing of coronal radiation in the accretion disk.  Since most of the
reprocessed radiation is emitted as thermal radiation, the higher amount of
reprocessing gives rise to higher Compton cooling rates.  Therefore, once
pair-production becomes important, an increase in the heating rate results
in a decrease in the temperature.  Fig.~2 of paper~I shows the allowed
parameter space of self-consistent corona temperatures and opacities.

\begin{figure*}
\centerline{\psfig{file=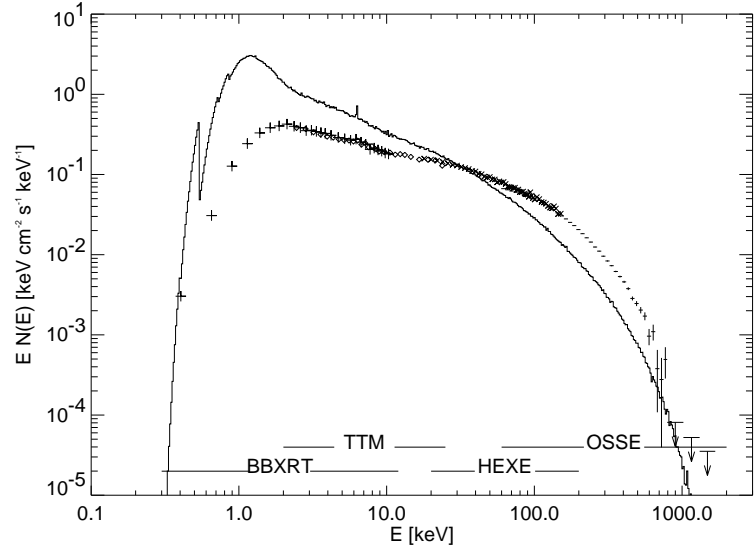,width=0.6\textwidth,angle=0}}
\caption{Slab geometry: Comparison of the predicted spectrum of escaping
radiation, modified by interstellar absorption, to the composite X-ray
spectrum of Cyg~X-1.  For our model, $\Tbb = 200$\,\eV, \tauT = 0.28, \Tc =
110\,\keV, and $N_{\rm H} = 6 \times
10^{21}$~cm$^{-2}$.}\protect\label{fig:cygslab}
\end{figure*}

In Fig.~\ref{fig:cygslab} we show our by-eye ``best-fit'' of the predicted
angle-averaged spectrum, modified by interstellar absorption, to the
composite spectrum of Cyg~X-1. 
For the slab geometry, the modeled spectrum
shown is the hardest spectrum possible while having a cut-off energy
$E_{\rm c} \gta 100$\,\keV. It is apparent, however, that the predicted
spectrum is still much softer than the observed spectrum.  In our grid of
models, the range of seed opacities is $0.1 \le \taup \le 2.0$ and the
range of coronal compactness parameters is $0.1 \le \lc \le 10^3$ (see
paper~I).  Out of the entire grid, the hardest spectrum predicted by our
self-consistent models has a photon index of $\alpha \sim 1.8$, while
Cyg~X-1 has a power-law index of $\sim 1.6-1.7$.  As discussed in
\cite*{wilms97a}, The BBXRT archived data were deconvolved by assuming a
power-law continuum, with a photon index of $\alpha = 1.62$, absorbed by
the interstellar medium with $N_{\rm H} = 6.0 \times 10^{21}$ cm$^{-2}$.
The TTM data were reduced by Borkous et al. (1995), the HEXE data by
\cite*{doeber95a}, and the OSSE data by \cite*{kurfess95a}.  Since the
models provided reasonable reduced chi-squared values ($\lta 1.5$), the
unfolded data should be at least a good approximation to the actual
spectrum.  Therefore, even though we are comparing our predicted spectra
with unfolded data, which are dependent on the assumed model, the result
that the predicted spectrum is always softer than the observed spectrum is
not sensitive to the deconvolution procedure.

According to our {\it linear} (and therefore not self-consistent) MC
simulations, simulated spectra with $\alpha \lta 1.7$,
while $\Tc \gta 100$\,\keV, are achieved only for models with $\tau \gta 0.3$.
These values are consistent with past work, in which the spectrum of
Cyg~X-1 has been described by ADC models with a slab geometry
(\cite{haardt93a,titarchuk94b}).  Such models, however, are not
self-consistent, for the temperature and opacity values lie within the
forbidden region of Fig.~2 of paper~I.  We remind the reader that the
predicted maximum temperatures are indeed upper limits, for our models do
{\it not} include additional cooling mechanisms such as bremsstrahlung.
(This mechanism was found to be negligible for the models of interest, but
its inclusion would reduce the maximum temperatures allowed.)

In addition to the simulated spectra having too soft a power-law component,
our models predict a very large Fe~K$\alpha$ equivalent width (EW).  Our
``best-fit'' slab model predicts an ${\rm EW} \sim 120$\,\eV, considerably
larger than the observed value of ${\rm EW} \lta 30$\,\eV\ 
(\cite{ebisawa96b}).  Finally, the slab models also predict an excess of
radiation for energies $E\lta 1$\,\keV\ (i.e., a ``soft excess'').  Since
the ``best-fit'' slab models are optically thin, most of the thermal
emission emitted by the disk escapes the system without interacting with
the corona. We have accounted for Galactic absorption by using the observed
column density of hydrogen, $N_{\rm H} = 6 \times 10^{21}$ cm$^{-2}$
(\cite{wilms97a}).  Clearly, this is not sufficient to ``hide'' the soft
excess for the 200\,\eV\ disk models. Models with a higher value of $N_{\rm
H}$ predict too low of a flux for energies below $1$\,\keV, so this
soft-excess problem cannot be solved by having having arbitrarily large
hydrogen column densities in the model.  We note, however, that the BBXRT
observation is unique since a soft-excess was not observed
(\cite{marshall93a}).  Since Cyg~X-1 usually has a soft-excess component
(e.g., \cite{balucinska91a,balucinska95a,ebisawa96b}), the discrepancy
between the predicted flux and the observed flux at low energies should not
be considered a serious problem by itself.  However, it does appear that
our ``best-fit'' slab geometry model predicts a higher soft-excess
component than observed by the {\em ASCA} GIS detector for November 11,
1993 (\cite{ebisawa96b}).



{\em We conclude that self-consistent accretion disk corona models with a
slab geometry are not capable of reproducing the observed broad X-ray
spectrum of Cyg X-1.}

\section{Sphere+Disk Geometry}
\subsection{The Model}\label{sec:sphmodel}
\begin{figure*}
\centerline{\psfig{file=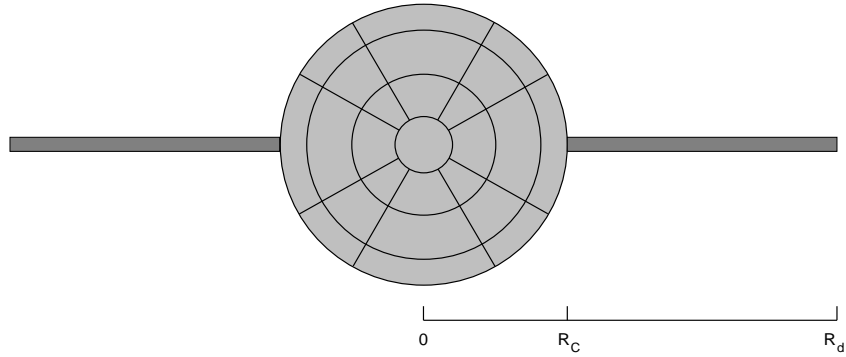,width=0.6\textwidth,angle=-90}}
\caption{The sphere$+$disk geometry. Spherical corona with exterior
accretion disk.  The corona has a radius $\Rc$, and the accretion disk
(assumed to be optically thick) has an inner radius $\Rc$ and an outer
radius $\Rd$.  The corona has been divided into several zones, such that
non-uniform temperature and opacity structures can be studied.  For
this paper, we assume uniform heating with volume within the corona.
\label{fig:sphere}}
\end{figure*}

The main limitation of the slab geometry models, in regard to explaining
the high-energy spectra of Cyg~X-1, is that there is too much reprocessing
of coronal radiation in the cold accretion disk.  Therefore, models with
the reprocessing matter having a smaller covering fraction (as seen from
the corona) are more likely to explain the observations.  One such geometry
is a combination of a spherical corona with an optically thick,
geometrically thin, cold accretion disk. Here, the accretion disk is
assumed to be exterior to the corona, as shown in
Fig.~\ref{fig:sphere}.  
This geometry is very similar to the
two-temperature accretion disk model of \cite*{shapiro76}, although, in our
model, the proton temperature is assumed to be equal to the electron
temperature.  This geometry is also similar to the advection-dominated disk
models by Narayan and collaborators (\cite{narayan95a,narayan95b};
\cite{narayan96a}, and references therein).  However, in our models, the
seed photons for Comptonization are produced by the accretion disk through
thermal emission rather than bremsstrahlung and synchotron radiation within
the corona.  The accretion disk extends from \Rc\ to \Rd\, where \Rc\ is
the radius of the sphere and \Rd\ is the outer radius of the disk.  We
define the ratio of the outer disk radius to the corona radius as $a =
\Rd/\Rc$.  For this paper, we set $a=10$, but the main results are not
sensitive to this value (see \S~\ref{sec:sphresults}).  

\subsubsection{Energetics}
We define the coronal compactness parameter by
\begin{multline}\label{eq:lcsphere}
\lc = \frac{\sigmaT}{m_{\rm e}c^3}\frac{\Lc}{\Rc} = \\ 4\pi
\left(\frac{m_{\rm p}}{m_{\rm e}}\right)\frac{\Lc}{L_{\rm edd}}\frac{R_{\rm
s}}{\Rc}
= 23\left(\frac{\Lc}{0.1L_{\rm edd}}\right)\left(\frac{100
R_{\rm s}}{\Rc}\right).
\end{multline}
Similar to the slab geometry models, we allow for intrinsic emission of
thermal radiation by the accretion disk (in addition to emission due to the
reprocessing of coronal radiation). We define $fP_{\rm G}$ to be the rate
at which gravitational energy is dissipated directly into the corona, where
$P_{\rm G}$ is the {\it total} rate of gravitational energy dissipated into
the system.  Consequently, $(1-f)P_{\rm G}$ is the rate of energy
dissipation into the disk.  The total luminosity of the accretion disk is
given by
\begin{equation}\label{eq:lumdisk}
L_{\rm d} = (1-f)P_{\rm G} + L_{\rm abs},
\end{equation}
where $L_{\rm abs}$ is the rate at which energy is absorbed (not
reflected) by the disk due to the reprocessing of radiation emitted by the
corona.  Equation~(\ref{eq:lumdisk}) can be expressed in terms of compactness
parameters as
\begin{equation}
\ld = (1-f)l_{\rm G} + l_{\rm abs},
\end{equation}
where $\ld = (\sigmaT/\me c^3)L_{\rm d}/\Rc$, $l_{\rm G} = (\sigmaT/\me
c^3)P_{\rm G}/\Rc$, and $l_{\rm abs} = (\sigmaT/\me c^3)L_{\rm abs}/\Rc$.
As with the slab geometry models, we constrain the models by setting
$(1-f)l_{\rm G} = 1$ for all of the simulations.  With this choice, $f$ is
given by
\begin{equation}
f = \frac{\lc}{1+\lc}.
\end{equation}
As discussed in paper~I, setting $(1-f)l_{\rm G}$ to unity allows us to
consider models where $0.01 \le f \le 1.0$, but models with other values of
$(1-f)l_{\rm G}$ yield the same ranges of self-consistent coronal
temperatures and opacities.

\subsubsection{Disk Temperature}
\begin{figure*}
\centerline{\psfig{file=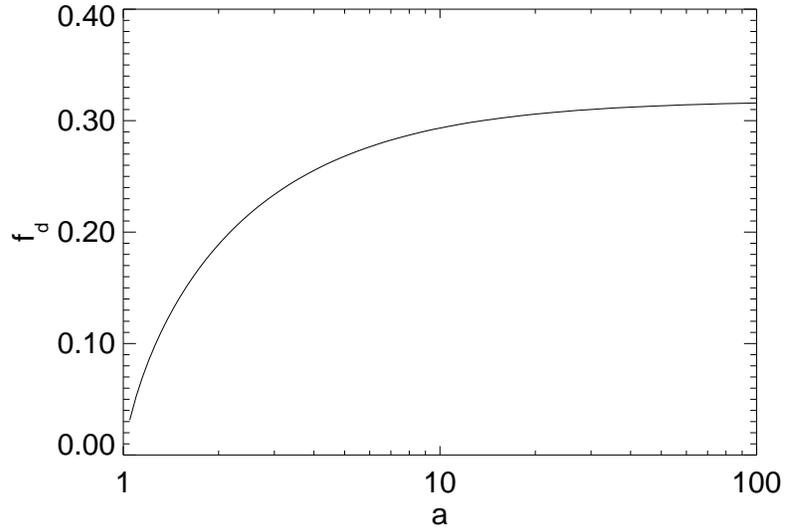,width=0.6\textwidth,angle=-0}}
\caption{The covering fraction, $\fdd(a)$, of the accretion disk with
an outer radius $a$ (normalized to the coronal radius), averaged
over the surface of the corona (spherical geometry).
\label{fig:fdisk}}
\end{figure*}

As in the slab-geometry model, we can estimate a lower limit to the disk
temperature by assuming that the disk is heated solely by
illumination from the corona;
\begin{equation}\label{eq:illumination}
\pi\Rc^2(a^2 - 1)\sigmasb \avg{\Tbb}^4 = \fdd(a) (1-A) \frac{\Lc}{2}, 
\end{equation}
where \Lc\ is the total luminosity leaving the corona and $\fdd(a)$\,is the
fraction of photons leaving the corona that is reprocessed in the disk.
This fraction can be approximated by assuming that the photons leaving the
corona are uniformly distributed over the surface of the corona with an
isotropic distribution.  With these assumptions, the fraction of escaping
photons that hits the disk is given by
\begin{eqnarray}\label{eq:fdisk}
\fdd(a) &=& \frac{\avg{\Delta\Omega}}{2\pi}\nonumber\\
  &=& \frac{1}{2\pi}\int_{\theta_{\rm
m}}^{\pi/2}\sin(\theta)\Delta\Omega(\theta){\rm d}\theta,
\end{eqnarray}
where $\avg{\Delta\Omega}$ is the average solid angle of the disk as seen
from the surface of the corona,
\begin{displaymath} \Delta\Omega(\theta) = 
\int_{\csc\theta}^{a}\frac{2\cos\theta}{x^2+1-2x\sin\theta}\left[\frac{a^2
-x^2}{a^2+1-2x\sin\theta}\right]^{1/2}{\rm d}x,
\end{displaymath}
$\theta_{\rm m} = \arcsin(1/a)$, and $\theta$ is the angle between the
normal of the coronal surface and the plane of the disk.
In Fig.~\ref{fig:fdisk}, we show how the covering fraction depends on $a$.
It is apparent that $f_d(a)$ is nearly constant for $a \gta 10$, and
therefore the amount of reprocessing of radiation within the accretion disk
is insensitive to $a$.  The maximum value of \fdd\ is $\approx 1/3$.  In
contrast, for a slab geometry, all of the downward directed radiation at
the base of the corona interacts with the accretion disk.  Therefore, the
slab geometry models predict much more prominent reprocessing features in
the spectrum of escaping radiation.

\begin{figure*}
\centerline{\psfig{file=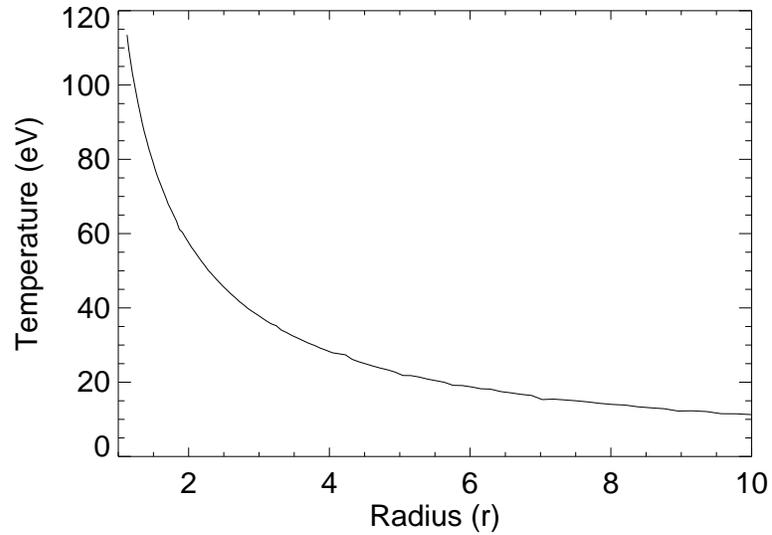,width=0.6\textwidth,angle=-0}}
\caption{The temperature of the accretion disk as a function of radius.
The temperature is calculated by assuming the disk is heated by
illumination of coronal radiation and cooled by thermal emission
(equation~(\ref{eq:TbbR})). Here, $\lc = 25$, $M = 10\Msun$, and $\Rc =
100\Rs$.\label{fig:TbbR}}
\end{figure*}

For large accretion disks (i.e., $a\gg 1$), the local disk temperature can
vary by more than an order of magnitude between the inner and outer radius.
Rather than using an average disk temperature for calculating
the spectrum emitted by the disk, a more proper treatment is to take into
account the temperature structure.  We can estimate the disk temperature as
a function of radius by equating the flux of absorbed coronal radiation
with thermal emission,
\begin{equation}
\sigmasb \Tbb^4(r)\,2\pi r\diff r =\frac{1}{2}(1-A)\Lc \diff\fdd,
\end{equation}
where $r = R/\Rc$, $\diff\fdd(r)/\diff r$ is the differential covering
fraction of a ring of radius $r$, and $\fdd(r)$ is determined numerically
using equation~(\ref{eq:fdisk}).  Using equation~(\ref{eq:lcsphere}), the
temperature can finally be expressed as
\begin{multline}\label{eq:TbbR}
\Tbb(r) = 150 \left(\frac{\diff\fdd(r)}{\diff r}\right)^{1/4}
\left(\frac{\lc}{25}\right)^{1/4} \\ \cdot \left(\frac{100
\Rs}{\Rc}\right)^{1/4} \left(\frac{10\Msun}{M}\right)^{1/4} {\rm eV}.
\end{multline}
In Fig.~\ref{fig:TbbR} we show how $\Tbb(r)$ varies with $r$.  From a
numerical fit, we find that $\Tbb(r)$ roughly decreases as $\Tbb(r)\propto
r^{-1.1}$ as compared to a power-law of $3/4$ corresponding to the standard
$\alpha$--disk models (\cite{shakura73}).
It should be remembered, however, that if any accretion-energy were
dissipated directly into the accretion disk ($f < 1$), then the disk
temperature would be higher than the values given above. To address this
possibility, we consider a simple model where, for $R>\Rc$, the accretion
disk behaves as the standard $\alpha$ disk (\cite{shakura73}), while, for
$R<\Rc$, all of the accretion energy is dissipated into the corona. In
order to determine the relative importance of the internal dissipation of
energy, compared to the contribution from coronal illumination, we estimate
the disk temperature by neglecting the contribution from illumination.  In
addition, we make the simplifying assumption that the disk is optically
thick and radiates as a blackbody such that the emitted flux
is equal to the dissipation rate,
\begin{equation}
\sigmasb \Tbb^4(R) = D(R) = \left(\frac{3GM\dot{M}}{8\pi R^3}\right)^{1/4},
\end{equation}
where we have assumed $R\gg\Rs$ (Frank, King, \& Raine, 1992).  Defining
$L_{\rm acc} =\eta 
\dot{M} c^2$, where $\eta$ is the efficiency by which the accretion process
converts gravitational energy into radiation, and using the standard
definition of the Eddington luminosity, $L_{\rm edd}$, we can express the
disk temperature as
\begin{multline}\label{eq:Talpha}
k\Tbb(R) = k\left(\frac{3}{16}\frac{m_{\rm p}
c^5}{GM\sigmasb\sigmaT}\right)^{1/4} \left(\frac{L_{\rm acc}}{\eta L_{\rm
edd}}\right)^{1/4}\left(\frac{R}{\Rs}\right)^{-3/4}\\ 
= 55\left(\frac{10\Msun}{M}\right)^{1/4}\left(\frac{L_{\rm acc}}{\eta L_{\rm
edd}}\right)^{1/4}\left(\frac{R}{100\Rs}\right)^{-3/4} {\rm eV}.
\end{multline}
By comparing equation~(\ref{eq:Talpha}) with equation~(\ref{eq:TbbR}) (or
Fig.~\ref{fig:TbbR}), it appears that the heating rate due to the direct
dissipation of accretion energy is comparable to that from coronal
illumination.  A proper treatment of the disk temperature structure would
be to equate thermal emission with both heating mechanisms.  Such a
treatment is outside the scope of this paper; with a heating rate equal to
twice the value used in equation~(\ref{eq:Talpha}), the disk temperature will be roughly $1.2$
times higher than the value given in equation~(\ref{eq:Talpha}) ($\Tbb^4$ is
proportional to the total heating rate per unit area).  Therefore, the
inclusion of direct energy dissipation into the disk changes the disk's
total luminosity, but does not significantly change 
the temperature profile of the disk.

For simplicity, we have assumed that the intrinsic flux of thermal
radiation is constant with respect to radius (the emission due to
reprocessing, however, is determined locally by equating emission with
absorption).  This approximation is not too important since the emission
due to reprocessing is higher than the emission due to internal dissipation
for the models considered in this paper.  In regards to the soft excess
issue, for $a \approx 10$, these models slightly overestimate the predicted
amount of soft excess by less than 10\%.

\subsubsection{Covering Fraction of Corona}
\begin{figure*}
\centerline{\psfig{file=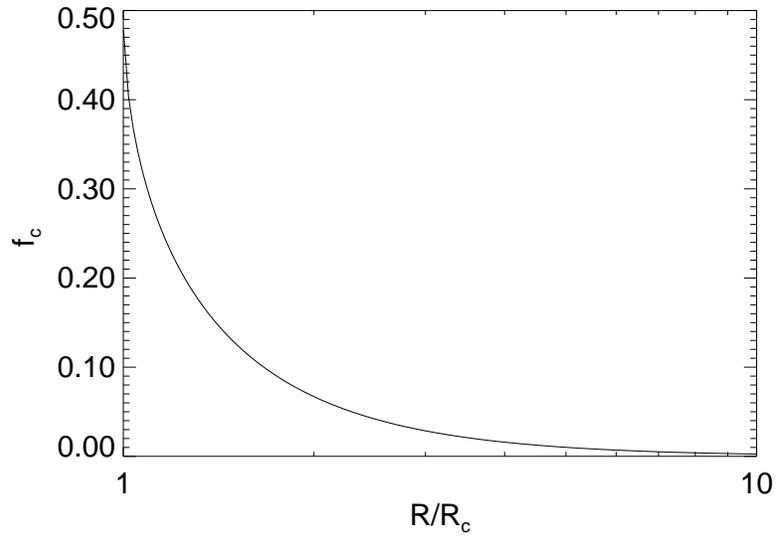,width=0.6\textwidth,angle=-0}}
\caption{The covering fraction of the spherical corona (with radius \Rc)
evaluated on the disk at a radius $R$. \label{fig:fcor}} 
\end{figure*}

\begin{figure*}
\centerline{\psfig{file=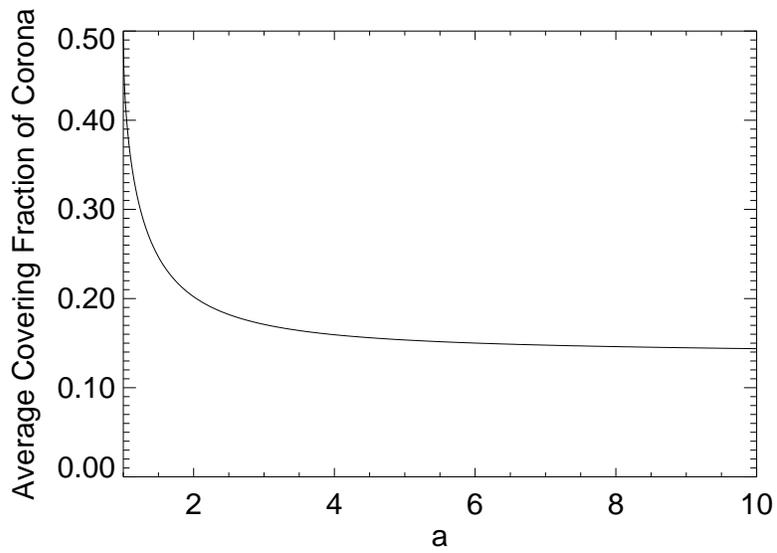,width=0.6\textwidth,angle=-0}}
\caption{The covering fraction of the spherical corona, averaged over the
accretion disk with outer radius $a=\Rd/\Rc$ and weighted by the flux of
thermal radiation, $F(r) \propto r^{-3.5}$.
\label{fig:avgfc}}
\end{figure*}

In all of our models, the seed-photons are produced within the cold
accretion disk (we neglect bremsstrahlung radiation).  For this geometry,
however, not all seed photons have to propagate through the corona prior to
escaping the system.  Therefore, even for optically thick models, a very
large soft excess of radiation is always predicted.  The covering fraction
of the corona, evaluated on the accretion disk at a distance $d$ from the
center, is given by
\begin{equation}
\fc(R) = \frac{\Delta\Omega}{2\pi} = \frac{1}{2}\left[1-
\frac{\sqrt{\left(\frac{\R}{\Rc}\right)^2-1}}
{R/\Rc}\right],
\end{equation}
where \Rc\ is the radius of the corona.  As shown in Fig.~\ref{fig:fcor},
the covering fraction of the corona rapidly decreases with increasing $R$.
In order to estimate the fraction of the thermal radiation
emitted by the accretion disk that interacts with the corona, we average \fc\
over the accretion disk, weighted by the thermal flux, $F(r)$,
\begin{equation}\label{eq:avgfc}
\avg{f_{\rm c}}(a) = \frac{\int_1^a
2\pi r \fc(r) F(r) dr}{\int_1^a 2\pi r
F(r) dr}.
\end{equation}
For an accretion disk heated solely by illumination, we found that
$T(r)\propto r^{-1.1}$ (Fig.~\ref{fig:TbbR}).  On the other hand, for the
case where the disk is heated solely by internal dissipation, i.e., the
standard $\alpha$-disk, $T(r)\propto r^{-3/4}$ (\cite{shakura73}). For a
rough estimate of $\avg{\fc}$ when both disk heating mechanisms are
important, we assume $F(r)=kT^4(r) \propto r^{-3.5}$. In
Fig.~\ref{fig:avgfc}\ we show how this averaged covering fraction depends
on $a$. 
For $a \gta 2$, $\avg{\fc} \lta 0.2$, and at least $80$\% of the
thermal radiation escapes the system without interacting with the corona,
even if the corona is optically thick.  In contrast, for the slab geometry
models, all of the thermal radiation must propagate through the corona
prior to escaping the system.  Therefore, the sphere$+$disk models always
predict that a large amount of thermal radiation, relative to the amount of
Comptonized radiation, will escape the system.  Unless this thermal
radiation is absorbed by the interstellar matter (ISM), it will appear as a
soft excess of radiation.

For $\tauT \gta 1$, the self-consistent temperature structure varies by
roughly 30\% for the case where the heating rate is uniform with volume.
For these optical depths, the outer shells of the corona are the hottest
while the center region is the coldest.  This variation is not large enough
such that the predicted spectra contain a ``hardening'' feature that mimics
the hardening due to Compton reflection (\cite{skibo95b}).

\begin{figure*}
\centerline{\psfig{file=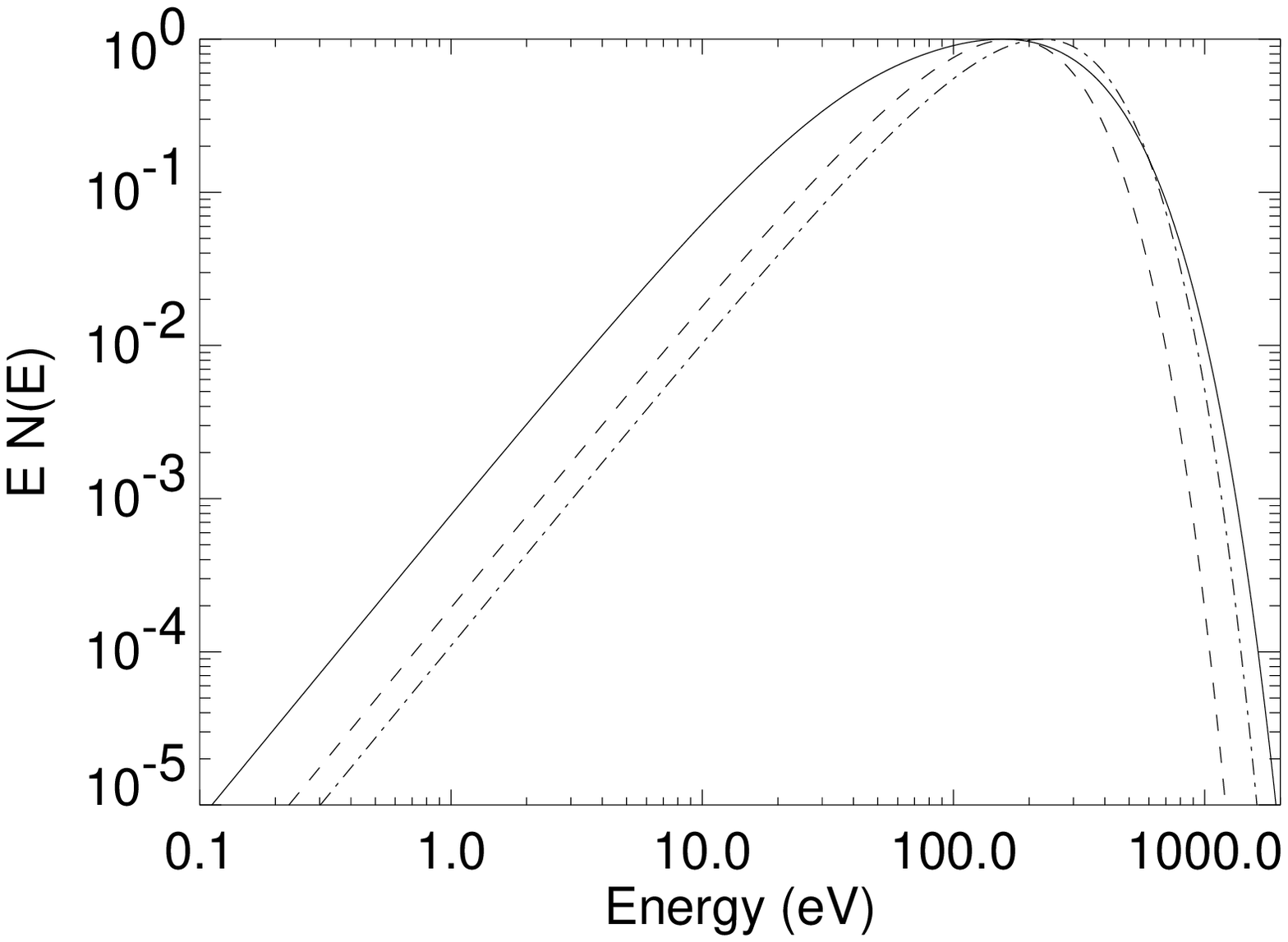,width=0.6\textwidth,angle=-0}}
\caption{The spectrum of thermal radiation emitted from an accretion
disk with a radial temperature structure, $\Tbb(r)$, as given by
equation~\ref{eq:TbbR} (solid line). Here, $\lc = 23$, $\Rc = 100 \Rs$, and
$M = 7 \Msun$.  Dashed line is a Planckian distribution corresponding to a
temperature $\Tbb = 60$\,\eV.  Dashed-dot-dashed line is a Planckian
distribution corresponding to a temperature $\Tbb = 80$\,\eV. All
distributions have been normalized to unity.
\label{fig:specbb}}
\end{figure*}

For the sphere$+$disk geometry, the accretion disk
receives less illumination from the corona since the covering fraction is
lower than in the slab geometry case, while the surface area of the disk is
larger.  In \S\ref{sec:sphmodel}, we estimated the temperature structure
due to either coronal illumination or internal dissipation of accretion
energy (equation~\ref{eq:Talpha}).  For either case, 
\begin{equation}
\Tbb(r) \propto r^{-\gamma},
\end{equation}
where $r = R/\Rs$, $\gamma \approx 1.1$ for heating by coronal illumination
and $\gamma = 0.75$ for the standard $\alpha$-disk.  The flux of radiation
emitted by an accretion disk is approximated by
\begin{equation}
F(E) \propto \int_1^aB_{\rm E}(T) r \diff r,
\end{equation}
where $B_{\rm E}(T)$ is the Planckian distribution.  In
Fig.~\ref{fig:specbb}, we show the corresponding spectrum, $F(E)$, for
$\gamma = 1.1$, and we compare $F(E)$ to a Planckian distribution
corresponding to the temperatures $\avg{\Tbb}=60$\,\eV\ and $80$\,\eV.  
It is seen that the Planckian distributions approximate the high-energy
tail fairly-well.  The disagreement at energies lower than $100$\,\eV\ is
irrelevant since this radiation is efficiently absorbed by the ISM.  In
addition, since the temperature decreases rapidly with radius, most of the
radiation having an energy $E\gta 100$\,\eV\ is emitted within the inner
region of the disk, and the exact value of $\gamma$ is unimportant.
Therefore, since the Comptonized spectrum is insensitive to the exact shape
of the spectrum of seed photons, we used a Planckian distribution
corresponding to a single disk temperature of $80$\,\eV\ to approximate the
exact integrated disk spectrum.

\subsection{Application to Cyg~X-1}\label{sec:sphresults}
\begin{figure*}
\centerline{\psfig{file=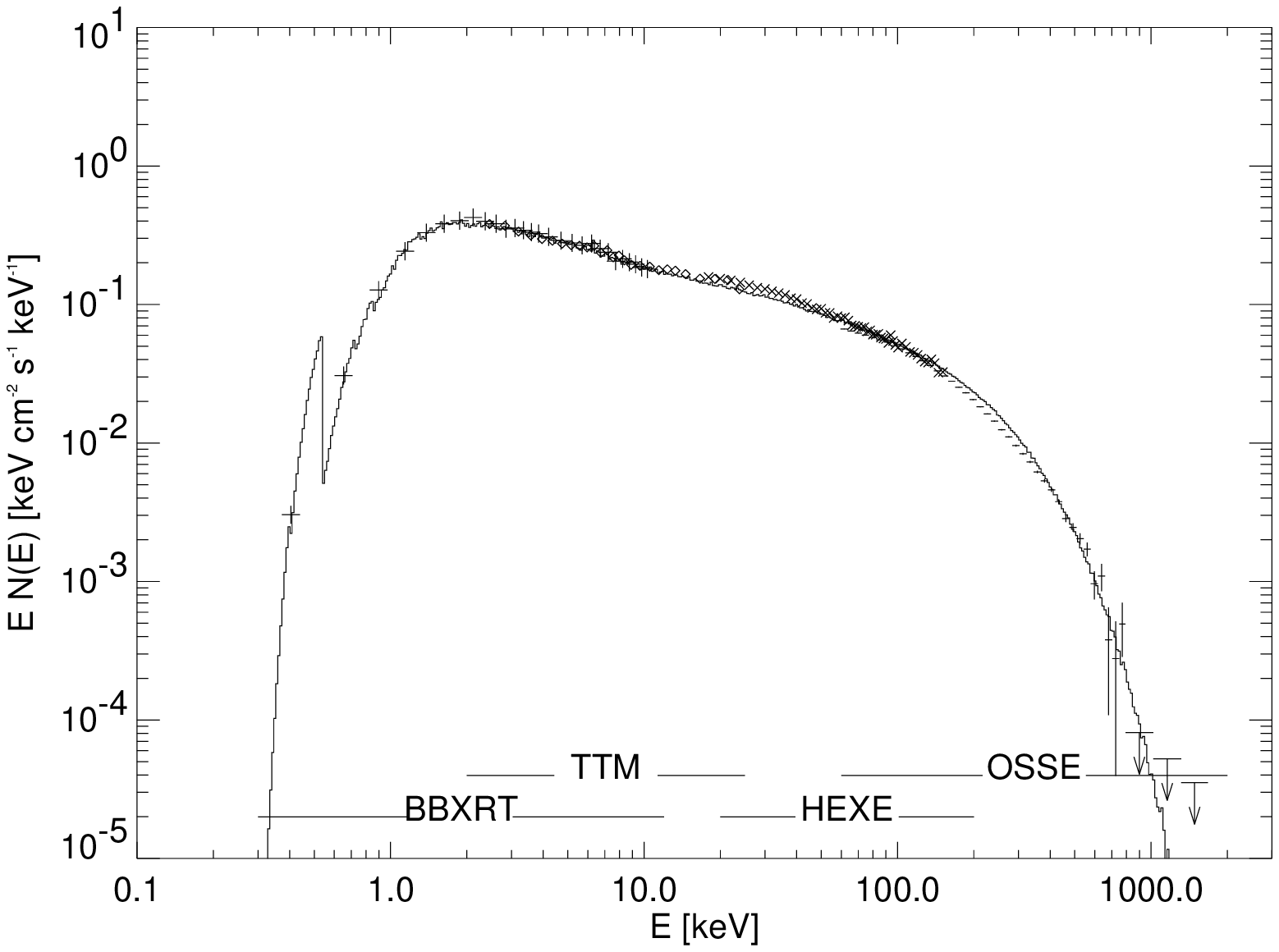,width=0.6\textwidth,angle=-0}}
\caption{Comparison of the predicted spectrum of escaping radiation,
modified by interstellar absorption, with the X-ray spectrum of Cyg~X-1.
This is for the sphere$+$disk geometry.  Here, $\Tbb = 50$\,\eV, \tauT =
1.5, \Tc = 90\,\keV, and $N_{\rm H} = 6 \times 10^{21}$~cm$^{-2}$.
\label{fig:cygsphb}}
%
\end{figure*}

We computed a grid of models, in which the range of seed opacities is $0.5
\le \taup \le 4.0$ and the range of coronal compactness parameters is $0.1
\le \lc \le 100$.  In Fig.~\ref{fig:cygsphb}, for $k\Tbb=50$\,\eV, we
compare the predicted spectrum of escaping radiation, modified by
interstellar absorption, of our ``best-fit'' model to the composite
spectrum of Cyg~X-1.  Since the seed photons enter the corona from the
exterior of the sphere, the effective optical depth of the sphere is
reduced.  
Therefore, the total optical depth (as defined by $\tauT =
\Rc\sigmaT(\nee+\np)$, where $\nee = \np + n_{+}$) must be higher than that
for the slab geometries in order for the models to predict similar
power-laws. Our ``best-fit'' model has a total optical depth $\tauT = 1.5$
and an average coronal temperature $\avg{\Tc} = 90$\,\keV.

\begin{figure*}
\centerline{\psfig{file=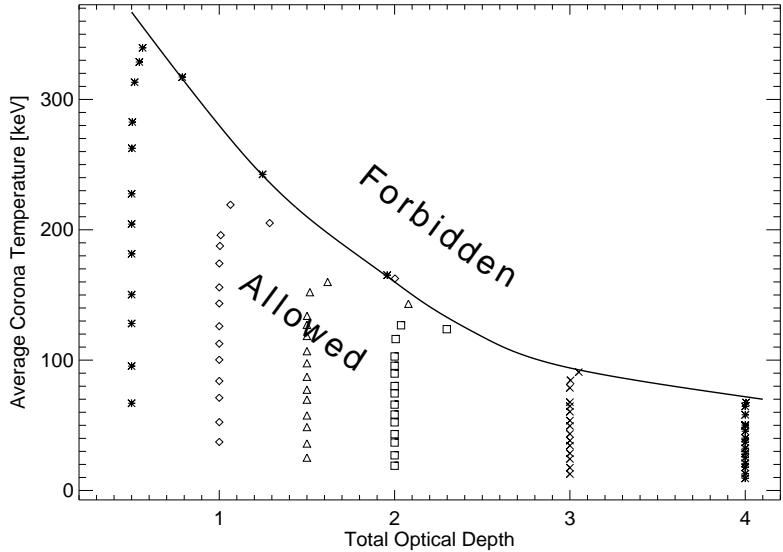,width=0.6\textwidth,angle=-0}}
\caption{Allowed temperature and opacity regime for self-consistent ADC
models with a sphere$+$disk geometry.  Solid line is derived from a ``fit
by eye'' to the numerical results.  For a given total optical depth,
temperatures above the solid line are not possible.  For all models, the
blackbody temperature of the disk is $k\Tbb = 50$\,\eV.  Different symbols
represent models with different seed opacities, while different points
having the same symbol represent different coronal compactness
parameters. \label{fig:ttausd}}
\end{figure*}

With this geometry, the corona is able to reach much higher temperatures as
compared to the slab models because the corona is ``photon-starved,'' i.e.,
the luminosity of the seed photons is much less than the luminosity of the
hard X-rays (\cite{zdziarski90a}).  As discussed above, most photons that
are reprocessed within the accretion disk do not re-enter the corona.
Therefore, the Compton cooling mechanism that prohibits slab geometry
coronae from having high temperatures is not as efficient in keeping the
spherical coronae cool.  In Fig.~\ref{fig:ttausd}, we show the allowed
regime of optical depths and average corona temperatures for the
sphere$+$disk geometry.  
Since the coronae are allowed to be much hotter
than the slab geometry models, there are self-consistent \{\Tc,\tauT\}
combinations such that the observed power-law and cut-off in BHCs can be
reproduced.  Since more than 90\% of the thermal radiation escapes the
system without passing through the corona, the soft excess is very large
for these models.  As in the case with the slab-geometry models, Galactic
absorption is not sufficient to ``hide'' the soft excess for the 200\,\eV\
disk models.  In Fig.~\ref{fig:cygsphb}, for $k\Tbb = 50$\,\eV, we compare
our best-fit model to the spectrum of Cyg~X-1.  Here, the predicted thermal
excess appears to be consistent with the observational data, and the
entire-broad band spectrum is adequately described by the model.  As
discussed in the Appendix, even though we are fitting `unfolded' data, we
are confident that this model provides a good representation of the
intrinsic photon-spectrum of Cyg~X-1.  However, the main point of this
paper is to motivate a more detailed study of the sphere+disk model, using
a more rigorous method of data analysis.

A nice feature of the sphere$+$disk model is that it naturally predicts an
``effective'' disk temperature to be much lower than in the slab geometry
models.  In fact, due to the lower disk temperature, this model appears to
predict a soft excess that is consistent with the observations
(\cite{balucinska91a,balucinska95a}). The quality of the observations,
however, is not too good for $E\lta 2$\,\keV, which is unfortunate since
this is the energy range where the spectrum is most sensitive to the disk
temperature.  Also, as noted in \S~\ref{sec:slabcyg}, Cyg~X-1 usually
contains a soft-excess component. 
We plan on comparing this model to better observational data in this
energy range (e.g., {\em RXTE} and {\em ASCA} data) in order to constrain
this model further.

To test whether the predicted thermal radiation from the accretion disk is
observable in the UV, we have compared the UV flux predicted by our model
for the spectral band from 1200\,\AA\ to 1950\,\AA\ with the flux observed
in this band by IUE. Our predicted flux is two to three orders of magnitude
smaller than the flux in an IUE Low Dispersion spectrum of HDE 226868 made
in June~1980. This is consistent with earlier findings that the UV spectrum
of the system HDE 226868 --- Cyg X-1 is dominated by emission from the O
star (\cite{treves80a,pravdo80a}). Therefore, IUE observations cannot be
used to constrain the disk temperature of our models.


Due to the smaller covering fraction of the accretion disk, the models
predict weaker reprocessing features in the escaping radiation field as
compared to the slab models.  In principle, in modeling BHCs, the value of
$a$ can be constrained by the equivalent width of the Fe~K$\alpha$
fluorescent line, which is proportional to the solid angle of the disk,
$\fdd(a)$.  For Cyg~X-1, the EW of the Fe~K$\alpha$ line is measured to be
${\rm EW} \lta 70$\,\eV\ (\cite{ebisawa96b,gierlinski97b}).
With $a = 10$ (corresponding to a disk covering fraction $\fdd \approx
0.3$), our best fit models predict an EW of $\sim 60$\,\eV.  Therefore,
models with smaller \fdd\ will predict EWs that are too low to be
consistent with the observations.  For $\fdd \gta 0.25$, $a \gta 3$
(Fig.~\ref{fig:fdisk}). 
On the other hand, since $\fdd(a)$ does not vary
significantly for $a \gta 10$, this method cannot provide an upper limit on
the size of the accretion disk.

\section{Discussion}
Our self-consistent slab accretion disk corona models are unable to explain
the broad-band X-ray spectra of Cyg~X-1. The modeled coronae are either too
cold or have an optical depth that is too small, resulting in spectra that
are much softer than the observed spectrum.  All previous ADC models that
have successfully described the spectra of Cyg~X-1 have temperature and
opacity values that are outside the allowed region of self-consistent
values.  In addition, as discussed in paper~I, the predicted angle-averaged
equivalent width (EW) of the Fe K~$\alpha$ fluorescent line is $EW \gta
150$\,\eV, a value roughly three times higher than the measured value
for Cyg~X-1 (\cite{ebisawa96b}), indicating that the covering fraction
of the reprocessing material is $\Omega/2\pi \sim 0.3$.

{\em We believe that there is no way around these shortcomings of the slab
geometry model and that these ADC models are not the appropriate models for
explaining the high-energy spectra of BHCs.}  This claim is in agreement
with \cite*{gierlinski97b}, who, by fitting the joint {\sl Ginga}-OSSE
observation of Cyg~X-1 with a power law plus a reflection component, find
that the solid angle of the cold medium is $\sim 0.3 \times 2\pi$,
significantly lower than the angle corresponding to the slab geometry, a
result consistent with \cite*{ebisawa96b}.  \cite*{gierlinski97b} also
argue that the corona must be ``soft-photon starved,'' which rules out a
slab geometry.

For ADC models with a sphere$+$disk geometry, there are self-consistent
temperature and opacity combinations such that the power-law and cut-off
portions of the spectra of BHCs can be explained.  In addition, due to the
small covering fraction of the disk as seen from the corona, it is possible
that the ``effective'' disk temperature (a flux-weighted average disk
temperature) can be as low as $\lta 80$\,\eV.  With such low values, the
models predict a thermal excess (or lack thereof) that appears to be
consistent with the observations of Cyg~X-1 since Galactic absorption is
more efficient at these lower energies.  It is still unclear, however,
whether an ADC model having this geometry is stable, as the
\cite*{shapiro76} model suffers from a thermal instability
(\cite{piran78a}).  In our models, we can only assume a corona/disk
morphology, but we cannot determine whether this morphology is thermally or
dynamically stable.  The sphere$+$disk geometry does approximate the
geometry corresponding to the {advection-dominated} models of
\citename{narayan96a} (\citeyear{narayan96a}, and references therein) which
appear to be thermally stable.  In addition, our best-fit models for
Cyg~X-1 have an optical depth $\tauT \approx 1.5$ and a coronal temperature
$k\Tc \approx 90$\,\keV, values that are consistent with the
advection-dominated models if the mass accretion rate $\dot{M} \sim 0.1 -
0.03 \dot{M}_{\rm edd}$, where $M_{\rm edd}$ is the Eddington accretion
rate (\cite{narayan96a}).

\begin{figure*}
\centerline{\psfig{file=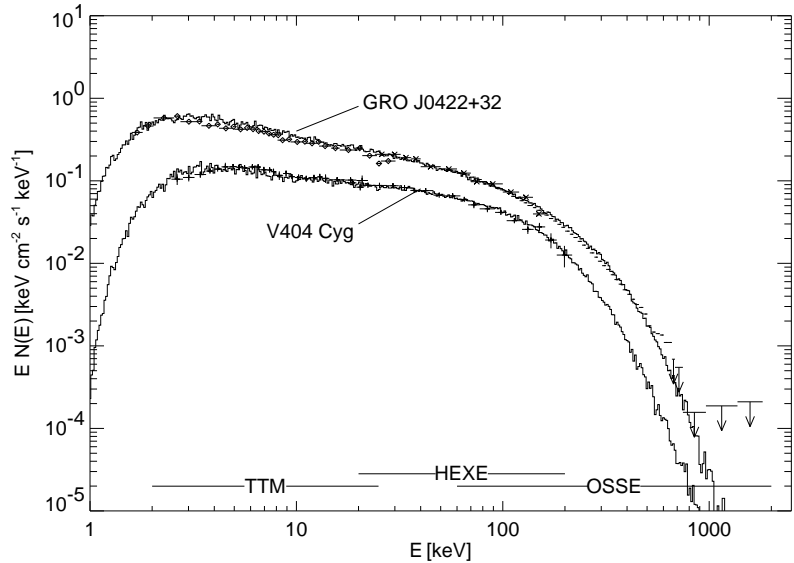,width=0.6\textwidth,angle=-0}}
\caption{Comparisons between the composite spectra of GRO J0422+32
(\protect\cite{kroeger96a}) and V404 Cyg
(\protect\cite{intzand93a,doeber94a}; June 20, 1989 spectrum) and the
predicted spectra from the best-fit sphere$+$disk models.  The spectra have
been scaled arbitrarily. For V404 Cyg, the parameter values of our best-fit
model are $k\Tbb = 52$\,\keV, $\tauT = 3.4$, and $N_{\rm H} = 2.4 \times
10^{22}$ cm$^2$.  For GRO J0422+32, $k\Tbb = 75$\,\keV, $\tauT = 1.9$, and
$N_{\rm H} = 1.6 \times 10^{22}\,{\rm cm}^2$.\label{fig:bhcs}}
\end{figure*}

The results obtained here for Cyg~X-1 are also applicable to other low
state BHCs.  In Fig.~\ref{fig:bhcs} we compare the spectra of the black
hole X-ray transients GS2023+338 and GRO J0422+32 to the predicted spectra
from our sphere+disk geometry model.  
Occasionally black hole X-ray
transients have spectra that are dominated by a strong soft-excess at lower
energies and a pure power-law component at high energies. In the low state,
however, GS2023+338 (=V404 Cyg) and GRO J0422+32 did not contain a soft
excess, and their spectra of are very similar to that of the low-state of
Cyg~X-1.  Therefore, it appears that photon starved ADC models, such as the
sphere$+$disk geometry, are applicable in explaining the observed X-ray
spectra of many BHCs.

Our ``fits'' to the unfolded data should be interpreted as a motivation to
consider this model in much more detail.  For example, we have recently
implemented our sphere+disk model in tabular form for the XSPEC
user-defined model.  We have recently implemented the sphere+disk geometry
spectral grid as an user-defined model for XSPEC (version 9,
\cite{arnaud96a}), and are currently analyzing new data of Cyg~X-1 and
other BHCs from the RXTE satellite. Our preliminary results are in
agreement with the results presented in this paper.



\section*{ACKNOWLEDGEMENTS}
We thank M.~Nowak for the many useful discussions. We also thank
K.~Ebisawa, the referee, for his useful suggestions which improved the
clarity of this paper. This work has been financed by NSF grants
AST91-20599, AST95-29175, INT95-13899, and NASA Grant NAG5-2026 (GRO Guest
Investigator program), DARA grant 50 OR 92054, and by a travel grant to
J.W.\ and M.G.M.\ from the DAAD.  This research has made use of data
obtained through the HEASARC Online Service, provided by NASA-GSFC, and of
data obtained through the IUE VILSPA data server, operated by the European
Space Agency.

\section*{Appendix A}
Comparisons of our models with observations have relied on the
photon-spectrum presented in Fig.~\ref{fig:cygsphb}. Although the use of
``unfolded'' X-ray spectra is very common in the astronomical literature,
it is important to keep in mind how these spectra are prepared. Due to the
poor energy resolution of today's X-ray detectors, the inversion of the
detector response-matrix is generally impossible (\cite{blisset79a}, and
references therein). Thus, in order to get an estimate of the intrinsic
photon spectrum from a source, a spectral model has to be fit to the
spectrum observed by the X-ray detector, usually using a $\chi^2$-reduction
method and an estimate for the detector response matrix (Gorenstein,
Gursky, \& Garmire, 1968).  If the $\chi^2$-value resulting from the
fitting process is small, then it is assumed that the modeled
photon-spectrum is a good representation of the observed
photon-spectrum. In order to produce an ``unfolded'' spectrum, the
additional assumption is made that the residual (for a given detector
energy channel) between the modeled spectrum (folded through the response
matrix) and the observed count-rate is a good representation of the
deviation between the spectral model and the intrinsic photon-spectrum (in
energy space).  Thus, at a given energy, the product of the modeled
photon-flux and the ratio between the observed count-rate and the
(estimated) model count-rate is what is called the ``unfolded'' spectrum.

Although a small $\chi^2$-value of the fit to the ``unfolded'' spectrum
usually implies that the model is an adequate description of the intrinsic
spectrum, sometimes this is not true.  Fitting our models to the observed
data in detector-space (using the method described above) would, in
principle, give the best method of determining the ``goodness of fit.''
However, the CPU time needed to obtain model spectra with a signal to noise
ratio that is suitable for spectral fitting is prohibitively high, and we
were forced to use a simplified approach; for a particular detector, we
simulated an observation using our modeled spectrum as the intrinsic
spectrum, the correct detector response matrix, background estimate, and
observational Poisson noise. This simulated spectrum was then fit with
several spectral models that are typically used to describe the X-ray
spectra of BHCs (e.g., a power-law with an exponential cutoff).  

We performed these tests using several typical X-ray detectors (ROSAT,
XTE-PCA, HEXE) and using typical count-rates and observation times for
these detectors. In all of the tests the fit-parameters found using our
model spectrum were in agreement with typical fit-parameters found in the
real observation of Cyg~X-1. In the simulated XTE-PCA observation, e.g.,
the model spectrum below 30\,\keV\ can be roughly described by an absorbed
power-law with an index of $\alpha\approx 1.62$. This index is similar to
those found, e.g., by \cite*{done92a} and \cite*{doeber95a}. In the
residuals of the simulation the Fe line and the reflection hump are clearly
visible. Using a more appropriate model that incorporates reflection, the
simulated spectrum is best described by reflection of a power-law of index
$\Gamma\approx 1.72$ off a cold accretion disk with covering factor 0.4.
Using an assumed length of 5\,ksec for the ROSAT-PSPC observation it is
possible to see evidence for the soft-excess in the simulated
observation. According to this simulated observation, the soft-excess as
seen by ROSAT PSPC can be described by a black-body with a temperature of
about 100\,eV or less, which is roughly comparable with the observations by
\cite*{balucinska95a}.  Therefore, we are confident that the model-spectrum
of Fig.~\ref{fig:cygsphb} is a good representation of the intrinsic
photon-spectrum of Cyg~X-1.


\begin{thebibliography}{}

\bibitem[\protect\astroncite{Aab et~al.}{1984}]{aab84a}
Aab, O.~{\'E}., Bychkova, L.~V., Kopylov, I.~M., {Kuma\u{\i}gorodskaya}, R.~N.,
   1984, Sov.\ Astron., 28, 90

\bibitem[\protect\astroncite{Arnaud}{1996}]{arnaud96a} Arnaud, K. A.,
1996, in: Jacoby, G. H., Barnes, J., eds, Astronomical Data Analysis
Software and Systems V, ASP Conf. Series Vol 101, San Francisco, p. 17

\bibitem[\protect\astroncite{{Ba\l{}uci\'{n}ska} \&  Hasinger}{1991}]{balucinska91a}
{Ba\l{}uci\'{n}ska}, M., Hasinger, G.,  1991, A\&A, 241, 439

\bibitem[\protect\astroncite{{Ba\l{}uci\'nska-Church}
  et~al.}{1995}]{balucinska95a}
{Ba\l{}uci\'nska-Church}, M., Belloni, T., Church, M.~J., Hasinger, G.,  1995,
  A\&A, 302, L5

\bibitem[\protect\astroncite{Blisset \& Cruise}{1979}]{blisset79a}
Blisset, R.~J., Cruise, A.~M.,  1979, MNRAS, 186, 45

\bibitem[\protect\astroncite{{Bochkar\"{e}v} et~al.}{1986}]{bochkarev86a}
{Bochkar\"{e}v}, N.~G., Karitskaya, E.~A., Luskutov, V.~M., Sokolov, V.~V.,
  1986, SvA, 30, 43

\bibitem[\protect\astroncite{{D\"obereiner} et~al.}{1995}]{doeber95a}
{D\"obereiner}, S., et~al., 1995, A\&A, 302, 115

\bibitem[\protect\astroncite{{D\"obereiner} et~al.}{1994}]{doeber94a}
{D\"obereiner}, S., et~al., 1994, A\&A, 287, 105

\bibitem[\protect\astroncite{Dolan}{1992}]{dolan92a}
Dolan, J.~F.,  1992, ApJ, 384, 249

\bibitem[\protect\astroncite{Done et~al.}{1992}]{done92a}
Done, C., Mulchaey, J.~S., Mushotzky, R.~F., Arnaud, K.~A.,  1992, ApJ, 395,
  275

\bibitem[\protect\astroncite{Dove et~al.}{1997}]{dove97a}
Dove, J.~B., Wilms, J., Begelman, M.~C.,  1997, ApJ,
\newblock submitted (paper I)

\bibitem[\protect\astroncite{Ebisawa et~al.}{1996}]{ebisawa96b}
Ebisawa, K., Ueda, Y., Inoue, H., Tanaka, Y., White, N.~E., 1996, ApJ, 467,
419 

\bibitem[\protect\astroncite{Frank et~al.}{1992}]{frank92}
Frank, J., King, A., Raine, D.,  1992,
\newblock Accretion Power in Astrophysics,
\newblock  (Cambridge: Cambridge University Press)

\bibitem[\protect\astroncite{George, Nandra, \& Fabian}{1990}]{george90}
George, I. M., Nandra, K., Fabian, A. C., 1990, MNRAS, 242, 28 com.

\bibitem[\protect\astroncite{Gierli\'nski et~al.}{1997}]{gierlinski97b}
Gierli\'nski, M., Zdziarski, A.~A., Done, C., Johnson, W. N., Ebisawa, K.,
Ueda, Y., Haardt, F., Phlips, B.~F., 1997, MNRAS,
\newblock in press

\bibitem[\protect\astroncite{Gies \& Bolton}{1982}]{gies82a}
Gies, D.~R., Bolton, C.~T.,  1982, ApJ, 260, 240

\bibitem[\protect\astroncite{Gies \& Bolton}{1986}]{gies86b}
Gies, D.~R., Bolton, C.~T.,  1986, ApJ, 304, 389

\bibitem[\protect\astroncite{Gorenstein et~al.}{1968}]{gorenstein68a}
Gorenstein, P., Gursky, H., Garmire, G.,  1968, ApJ, 153, 885

\bibitem[\protect\astroncite{Grebenev et~al.}{1993}]{grebenev93}
Grebenev, S., et~al., 1993, A\&AS, 97, 281

\bibitem[\protect\astroncite{Haardt}{1993}]{haardt93b}
Haardt, F.,  1993, ApJ, 413, 680,
\newblock (H93)

\bibitem[\protect\astroncite{Haardt et~al.}{1993}]{haardt93a}
Haardt, F., Done, C., Matt, G., Fabian, A.~C.,  1993, ApJ, 411, L95

\bibitem[\protect\astroncite{Haardt \& Maraschi}{1993}]{haardt93c}
Haardt, F., Maraschi, L.,  1993, ApJ, 413, 507,
\newblock (HM93)

\bibitem[\protect\astroncite{Herrero et~al.}{1995}]{herrero95a}
Herrero, A., Kudritzki, R.~P., Gabler, R., Vilchez, J. M., Gabler, A.,
1995, A\&A, 297, 556 

\bibitem[\protect\astroncite{Hutchings}{1978}]{hutchings78a}
Hutchings, J.~B.,  1978, ApJ, 226, 264

\bibitem[\protect\astroncite{Inoue}{1989}]{inoue89a}
Inoue, H., 1989, in Proc. 23rd ESLAB Symp. on Two Topics in X-ray Astronomy, 
Vol. 2, p. 783, ESA

\bibitem[\protect\astroncite{{in't Zand} et~al.}{1993}]{intzand93a}
{in't Zand}, J. J.~M., Pan, H.~C., Bleeker, J. A.~M., Skinner, G.~K.,
Gilfanov, M.~R., Sunyaev, R.~A., 1993, A\&A, 266, 283

\bibitem[\protect\astroncite{Kroeger et~al.}{1996}]{kroeger96a}
Kroeger, R.~A., et~al., 1996, A\&AS, 120, C117

\bibitem[\protect\astroncite{Kitamoto  et~al.}{1990}]{kitamoto90a}
Kitamoto, S., et~al., 1990, PASJ, 42, 85

\bibitem[\protect\astroncite{Kurfess}{1995}]{kurfess95a}
Kurfess, J. D., 1995, Adv. Space Res., 15(5), 103

\bibitem[\protect\astroncite{Liang \& Nolan}{1984}]{liang84}
Liang, E.~P., Nolan, P.~L.,  1984, Space Sci.\ Rev., 38, 353

\bibitem[\protect\astroncite{Marshall et~al.}{1993}]{marshall93a}
Marshall, F.~E., Mushotzky, R.~F., Petre, R., Serlemitsos, P.~J.,  1993, ApJ,
  419, 301

\bibitem[\protect\astroncite{Miyamoto et~al.}{1992}]{miyamoto92}
Miyamoto, S., et~al., 1992, ApJ, 391, L21

\bibitem[\protect\astroncite{Narayan}{1996}]{narayan96a}
Narayan, R.,  1996, ApJ, 462, 136

\bibitem[\protect\astroncite{Narayan \& Yi}{1995a}]{narayan95a}
Narayan, R., Yi, I.,  1995a, ApJ, 444, 231

\bibitem[\protect\astroncite{Narayan \& Yi}{1995b}]{narayan95b}
Narayan, R., Yi, I.,  1995b, ApJ, 452, 710

\bibitem[\protect\astroncite{Ninkov et~al.}{1987}]{ninkov87a}
Ninkov, Z., Walker, G. A.~H., Yang, S.,  1987, ApJ, 321, 425

\bibitem[\protect\astroncite{Nowak et~al.}{1996}]{nowak97a}
Nowak, M.~A., Vaughan, B.~A., Dove, J., Wilms, J.,  1996,
\newblock in Accretion  Phenomena and Related Outflows, IAU Coll.~163,
ed. D. Wickramasingle, L. Ferrario, G. Bicknell, 
\newblock in press

\bibitem[\protect\astroncite{Oda}{1977}]{oda77a}
Oda, M.,  1977, Space Sci.\ Rev., 20, 757

\bibitem[\protect\astroncite{Piran}{1978}]{piran78a}
Piran, T.,  1978, ApJ, 221, 652

\bibitem[\protect\astroncite{Pravdo et~al.}{1980}]{pravdo80a}
Pravdo, S.~H., White, N.~E., Kondo, Y., Becker, R.~H., Boldt, E.~A., Holt,
S.~S., Serlemitsos, P.~J., McCluskey, G.~E., 1980, ApJ, 237, L71

\bibitem[\protect\astroncite{Shakura \& Sunyaev}{1973}]{shakura73}
Shakura, N.~I., Sunyaev, R.~A.,  1973, A\&A, 24, 337

\bibitem[\protect\astroncite{Shapiro \& Lightman}{1976}]{shapiro76}
Shapiro, S.~L., Lightman, A.~P.,  1976, ApJ, 204, 187

\bibitem[\protect\astroncite{Skibo \& Dermer}{1995}]{skibo95b}
Skibo, J.~G., Dermer, C.~D.,  1995, ApJ, 455, L25

\bibitem[\protect\astroncite{Sokolov}{1987}]{sokolov87a}
Sokolov, V.~V.,  1987, Sov.\ Astron., 31, 419

\bibitem[\protect\astroncite{Stern et~al.}{1995}]{stern95a}
Stern, B., Begelman, M.~C., Sikora, M., Svensson, R.,  1995, MNRAS, 272, 291

\bibitem[\protect\astroncite{Sunyaev et~al.}{1993}]{sunyaev93a}
Sunyaev, R.~A., 1993, A\&A, 280, L1

\bibitem[\protect\astroncite{Sunyaev \& Titarchuk}{1980}]{sunyaev80}
Sunyaev, R.~A., Titarchuk, L.~G.,  1980, A\&A, 86, 121

\bibitem[\protect\astroncite{Tanaka \& Lewin}{1995}]{tanaka95a}
Tanaka, Y., Lewin, W. H.~G.,  1995,
\newblock in {X}-Ray Binaries, ed.
W.~H.~G. Lewin, J. {van Paradijs}, E.~P.~J. {van den Heuvel}
(Cambridge: Cambridge Univ.\ Press),  126

\bibitem[\protect\astroncite{Titarchuk}{1994}]{titarchuk94b}
Titarchuk, L.,  1994, ApJ, 434, 570

\bibitem[\protect\astroncite{Treves et~al.}{1980}]{treves80a}
Treves, A., et~al., 1980, ApJ, 242, 1114

\bibitem[\protect\astroncite{Wilms et~al.}{1997}]{wilms97a}
Wilms, J., Dove, J.~B., Maisack, M., Staubert, R.,  1997, A\&AS, 120, C159

\bibitem[\protect\astroncite{Zdziarski et~al.}{1990}]{zdziarski90a}
Zdziarski, A.~A., Coppi, P.~S., Lamb, D.~Q.,  1990, ApJ, 357, 149

\end{thebibliography}
\end{document}